\documentstyle[aps,prl,multicol,epsf]{revtex}

\begin{document}

\title {Emergence of Order in Textured Patterns}

\author{Gemunu H. Gunaratne$^{1,2}$, Anuradha Ratnaweera$^{2}$ and
        K. Tennekone$^{2}$}

\address{$^{1}$ Department of Physics,
               The University of Houston,
               Houston, TX 77204}
\address{$^{2}$ The Institute of Fundamental Studies,
                Kandy, Sri Lanka}
\maketitle
\nobreak

\begin{abstract}
A characterization of textured patterns,
referred to as the disorder function $\bar\delta(\beta)$, is used to 
study properties of patterns generated in the Swift-Hohenberg equation (SHE). 
It is shown to be an intensive, configuration-independent measure. 
The evolution of random initial states under the SHE 
exhibits two stages of relaxation. The initial phase,
where local striped domains emerge from a noisy background,
is quantified by a power law decay 
$\bar\delta(\beta) \sim t^{-\frac{1}{2} \beta}$. Beyond a sharp
transition a slower power law decay of $\bar\delta(\beta)$, which corresponds
to the coarsening of striped domains, is observed. 
The transition between the phases advances as the system
is driven further from the onset of patterns, and
suitable scaling of time and $\bar\delta(\beta)$ leads to the 
collapse of distinct curves.

The decay of $\bar\delta(\beta)$ during the initial phase remains 
unchanged when nonvariational terms are added to the underlying equations, 
suggesting the possibility of observing it in experimental systems.
In contrast, the rate of relaxation during domain coarsening
increases with the coefficient of the nonvariational term.

\end{abstract}

\pacs{PACS number(s): 05.70.Ln, 82.40.Ck, 47.54.+r}
\nobreak
\begin{multicols}{2}
\section{Introduction}

The study of spatio-temporal patterns has received considerable
impetus from a series of elegant experiments and theoretical developments
based on symmetry considerations. Recent experimental studies include 
those on reaction diffusion chemical systems \cite{ouyAswi1},
convection in fluids \cite{heuAgol} and gases \cite{bodAdeb},
ferrofluids \cite{rose}, and vibrated layers of granular 
material \cite{melAumb}. These results have been supplemented
with patterns generated in (relatively) simple model 
systems \cite{croAhoh,jacAmal,best}. The most complete theoretical treatments
of patterns rely on the study of symmetries of the underlying
system and those of the patterns \cite{golAste}. Unfortunately,
this analysis is restricted to periodic or quasi-periodic patterns.
A theoretical analysis of more complex states requires
the identification of suitable ``variables" to describe a given pattern.
Examples of such variables include the structure factor \cite{eldAvin},
the correlation length \cite{ouyAswi2,croAmei,chrAbra} and the 
density of topological defects \cite{houAsas}.
In this paper we study properties of another characterization,
referred to as the ``disorder function" \cite{gunAjon,gunAhof}.

The patterns studied are generated in physical systems 
(and models) whose control parameters are uniform in space and time;
thus, they result from spontaneous symmetry breaking. 
The simplest class of nontrivial structures are periodic.
They are typically striped, square, triangular or hexagonal
patterns that form in perfect, extended arrays \cite{croAhoh}. 
To obtain periodic patterns, the initial state of the system 
and/or the boundary conditions need to be carefully prepared. 
A second class consist of periodic patterns whose ``unit cells" 
have additional structure \cite{kudApie,judAsil}. A field describing periodic 
arrays can be expanded in a few plane waves. 

The patterns described above contain a unit cell that is repeated 
on a ``Bravais lattice" to generate a plane-filling 
structure. The qualitative description of the pattern
involves the characterization (in terms of symmetries) of the unit cell
and the generators of the Bravais lattice. For example, the
unit cell of a honeycomb lattice is $D_6$-symmetric, and the 
Bravais lattice is generated by two unit vectors $120^o$
apart.

Quasi-periodic patterns have also been observed under suitable experimental
conditions \cite{edwAfau}. Their symmetries can be observed
in Fourier space. For example, the spectrum of a quasi-crystal is 10-fold
symmetric \cite{levAste}. Quasi-periodic patterns can be described using a few 
``principal" plane waves along with their nonlinear couplings.

The bifurcations to and from a given periodic (or quasi-periodic) 
state can be studied using the ``Landau equations," which once again 
rely on the symmetries of the physical system and the 
pattern \cite{lanAlif}. The information used is that,
since the pattern is generated by symmetry breaking, a second 
pattern obtained under the action of any symmetry of the {\it physical
system} has identical features. The imposition of this equivalence 
(supplemented by the elimination of ``higher order" terms) gives the 
normal form equations of the pattern. They contain information on
aspects of dynamics of the pattern and details about its 
bifurcations \cite{golAste}.

Patterns such as those of Figs. \ref{set1} and \ref{set2} (which are generated 
in a model system) do not belong to the classes discussed above. These 
structures, referred to as ``textured" or ``natural" patterns \cite{cros}, 
are observed when the initial states from which they evolve
are not controlled. Similar structures are seen in small aspect
ratio systems when the boundaries play a significant role in the
creation of the pattern \cite{croAhoh}.

There is no (nontrivial) global symmetry of textures; 
consequently, they cannot be characterized using symmetry groups.
Note also that a second realization of the experiment (e.g.,
starting from a different set of initial conditions) will give 
a pattern that is vastly different in detail (such as 
Figs. \ref{set1}(a) and \ref{set1}(b)).
In spite of these differences, one can clearly
recognize similarities between distinct patterns. For example, the 
correlation length and the density of topological defects of the two textures
shown in Fig. \ref{set1} are similar. In contrast patterns generated under
other external conditions (e.g., Fig. \ref{set2}) have different 
characteristics. A theoretical treatment of textured patterns requires a
``configuration independent" description.

\begin{figure}
\narrowtext
\caption {Two patterns generated by evolving random initial states
via the Swift-Hohenberg equation for 1600 time units. The parameters used were 
$D=0.1$, $\epsilon=0.2$, $\nu=2$ and $k_0=1$. The initial states 
consisted of white noise
whose intensity varied between 0 and $10^{-3}$. Periodic boundary conditions 
were imposed on the square domain of $256\times 256$ lattice points, the length
of whose sides are $(48\pi / k_0)$.}
\label{set1}
\end{figure}

In Section II, we introduce the disorder function $\bar\delta(\beta)$
whose definition was motivated in part by the argument leading to the 
derivation of Landau equations; for patterns generated in uniform, 
extended systems, $\bar\delta(\beta)$ is required to be invariant 
under rigid motions of a labyrinthine pattern \cite{gunAjon,gunAhof}.
In section III, we briefly describe the method for evaluating the disorder
function from (typically noisy) grid-values of the field.
It relies on a method to approximate a continuous
function from values given on a grid referred to as the method of
``Distributed Approximating Functionals" (DAFs) \cite{dafref1}.

The main results of the paper, which include properties of the
disorder function, and its application to provide a 
quantitative description
of the relaxation of the patterns from an initially random state
are presented in Sections IV and V. The underlying spatio-temporal
dynamics is given by the Swift-Hohenberg equation (SHE) and one
of its variants. We first provide evidence to support the claim that the
disorder function consists of intensive, configuration independent
variables. The use of $\bar\delta(\beta)$ shows that pattern relaxation 
occurs in two distinct phases separated by a sharp
transition. We also study changes in the relaxation profile when the system
is driven further away from the onset of patterns. In Section V, we discuss
the effects of adding nonvariational terms to the SHE.

\begin{figure}
\narrowtext
\caption {Two patterns generated by evolving a random initial state
via the Swift-Hohenberg equation for 2400 time units. The parameters 
used for the integration were $D=0.01$, $\epsilon=0.4$, $\nu=2$ and $k_0=1/3$.
The initial states consisted of white noise
whose intensity varied between $\pm 10^{-2}$. The length of each side of
the square is $(48\pi / k_0)$.}
\label{set2}
\end{figure} 

\section{The Disorder Function}

Textured patterns observed in experimental 
systems \cite{ouyAswi1,heuAgol,bodAdeb,rose,melAumb} 
and those shown in Figs. \ref{set1} and \ref{set2} can be 
described by a scalar field $v({\bf x})$ which is smooth, except perhaps
at the defect cores. However, unless the patterns are trivial
(e.g., perfect stripes, target patterns) the analytical form of the 
field is unknown. Consequently, it is difficult to determine a set of 
``configuration independent" characteristics of structures
generated under similar conditions. We instead impose
a weaker requirement, that the characterizations remain invariant
under the action of the symmetries of the underlying physical
system; i.e., translations, rotations and reflections \cite{gunAjon}. 
Rather surprisingly, the measures so defined have similar values 
for distinct patterns such as those shown in Fig. \ref{set1}.

The most significant feature of labyrinthine patterns
is that they are locally striped; in a suitable neighborhood
$v({\bf x}) \sim sin ({\bf k\cdot x})$, where the modulus 
$k_0 (\equiv |{\bf k}|) $ of the
wave vector does not vary significantly over the pattern.
Structures generated in experiments and model systems include
higher harmonics due to the presence of nonlinearities in 
the underlying system; they only contribute to the
shape of the cross section of stripes. In order to use the
simplest characterization of textures, we eliminate the second and higher
order harmonics by the use of a suitable window function in Fourier space.
For experimental patterns (which do not have periodic boundary
conditions) this is a nontrivial task, and a method to
implement it is described in Ref. \cite{hofAgun}.

The simplest local field that is derived from $v({\bf x})$ and
whose value remains the same under all rigid motions is its Laplacian 
$\triangle v({\bf x})$. Terms such as $\triangle^n v^m({\bf x})$, 
though invariant, are difficult to extract from an incompletely 
sampled field (typically given on a square lattice).
The requirement that perfect stripes be assigned a null measure 
(they are not disordered), coupled with the local sinusoidal form of the 
(filtered) pattern implies that the lowest-order field relevant for our purpose
is $(\triangle + k_0^2) v({\bf x})$. The family of measures, 
referred to as the disorder function, is defined by
\begin{equation}
\delta(\beta) = (2-\beta){ {\int da |(\triangle + k_0^2) v({\bf x})|^{\beta}}
                 \over {k_0^{2\beta} <|v({\bf x})|>^{\beta}} },
\label{defn}
\end{equation}
where $<|v({\bf x})|>$ denotes the mean of $|v({\bf x})|$, and $\delta(\beta)$
has been normalized so that the ``intensive variables"
${ {\bar \delta}({\beta}) = \delta(\beta) / \int da }$ are scale invariant.
The moment $\beta$ is restricted to lie between 0 and 2 for reasons 
discussed below.
Local deviations of the patterns from stripes (due to curvature of
the contour lines \cite{gunAjon}) contribute to $\delta(\beta)$ through 
the Laplacian, while variations of the width of the stripes contribute 
via the choice of a ``global" $k_0$.

\begin{figure}
\narrowtext
\epsfxsize=3.0truein
\hskip 0.05truein
\epsffile{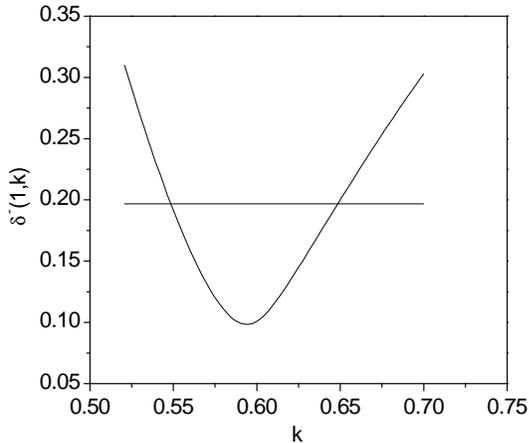}
\caption{The behavior of $\bar\delta(1)$ as a function of the wave-number
$k$ for a labyrinthine pattern. $k_0$ is estimated to be 
the (unique) minimum of the curve. The width $\Delta k$
is defined to be the distance between the $k$-values at which
$\bar\delta(1)$ reaches twice its minimum.}
\label{dvsk}
\end{figure}

$\bar\delta(\beta)$ depends on the choice of the wave-numer $k_0$
of the ``basic" stripes. Analysis of striped patterns 
$u_{st} ({\bf x}) = A sin ({\bf k\cdot x})$ and target patterns
$u_t ({\bf x}) = A cos(k r)$ show that $\bar\delta(1)$ is 
minimized when $k_0 = k = |{\bf k}|$. Studies of textured patterns from 
model equations indicate (see Fig. \ref{dvsk}) the presence of a unique 
minimum of $\bar\delta(1)$. We use the minimization of
$\bar\delta(1)$ as the criterion for the choice of the
wave-number $k_0$ in Eqn. (\ref{defn}).
For patterns generated using the Swift-Hohenberg equation \cite{croAhoh}, 
$k_0$ is very close to the wave-number obtained by 
minimizing the ``energy" \cite{pomAman}. We also find that our 
estimation of $k_0$ is far more robust (i.e., smaller variation
between distinct patterns) than that evaluated from the
power spectrum. This is presumably because wavelength variations
and curvature of contour lines at each location of the pattern
contribute to the computation of $\bar\delta(\beta)$. 

The variation in $k$ over the pattern (traditionally defined to be the
half width of the Structure Factor \cite{tuAcro}) 
can be estimated using the variation of
$\bar\delta(1)$ with the wave-number $k$ (see Fig. \ref{dvsk}). In the
remainder of the paper we define $\Delta k$ to be 
the distance between $k$-values for which $\bar\delta(1)$
is twice the minimum value \cite{footnote2}. 
Analysis of textures shows that $\Delta k$ is 
a configuration-independent, intensive variable.

Observe that our choice of $k_0$ is arbitrary in one sense; we could 
have chosen to minimize $\bar\delta(\beta)$
for any fixed $\beta$ to determine $k_0$. 
However, the observed variations in $k_0$ are insignificant.
Alternatively, we could have evaluated {\it each} $\bar\delta(\beta)$
by minimizing it with respect to $k$. We choose not to implement
this scheme because of the need to estimate only one free
parameter (i.e., the wave-number) for a given pattern.

For a perfect set of stripes the function $\bar\delta(\beta)= 0.$ 
A domain wall contains curvature of the
contour lines and variations of the stripe width; consequently
it has non-zero disorder. $\delta(\beta)$
for a single domain wall is a monotonically increasing function of the 
angle $\theta$ between the stripes of the two domains \cite{jone}.
Thus $\delta(\beta)$ provides information absent in
characterizations such as the correlation length. The disorder function
for a target pattern $v({\bf x}) = a cos(k_0 r) $ is  
known \cite{gunAjon}, and is used to determine 
the accuracy of the numerical algorithms.
For target patterns, the integral in the numerator diverges as
$(2-\beta)^{-1}$, and leads to limiting the range of $\beta (<2)$, and to the 
introduction of the prefactor in the definition of $\delta(\beta)$. 

The weights of distinct characteristics of a texture (e.g., domain walls, 
defects, variations of the stripe-width, etc.) depend on
the moment $\beta$. In particular, the contribution
to $\delta(\beta)$ from a domain wall vanishes as 
$\beta \rightarrow 2$, and the limit is proportional to the number
of targets in a pattern. The effects
of noise on the calculations are minimal. For example, 
addition of $10\%$ white noise typically changes $\delta(\beta)$ by less than 
$2\%$. 

\section{Distributed Approximating Functionals}

The critical requirement for a good estimate of $\bar\delta(\beta)$ 
is a sufficiently accurate determination of
the Laplacian. Calculation of derivatives of a field from values given on 
a lattice is a delicate task, especially in the presence of noise. 
A technique, utilizing what are referred to as Distributed Approximating
Functionals (DAFs), has
been introduced recently, to fit analytically or approximate
a continuous function from known values on a discrete
grid \cite{dafref1}. Unlike typical finite difference schemes, it estimates
the function and its derivatives using a range of neighboring
points ($\sim$ 40 in our case); consequently, the required 
computations are much less sensitive to noise.

The most useful for our application has been a class of DAFs for which
the order of accuracy of the fit is the same both on and off the
grid. (This is in contrast to interpolation, which forces the fit to be
exact on the grid, but always leads to intertwining about the function
off the grid, thus leading to less accurate estimation of derivatives.)
Alternatively, the method is designed so that there are no special
points. If the labels on the grid points are erased after DAF-fitting
of a function, it becomes essentially impossible to identify the
points that were on the grid. The most general derivation
of the DAFs is via a variational principle \cite{dafref2},
yielding
\begin{equation}
g_{DAF}(x)=\sum_{k}I(x,x_{k})g(x_{k}),
\label{daf0}
\end{equation}
where $k$ labels the grid points, $g(x_k)$ are the known input values
of the function (which may contain noise), and the sum is over
$x_k$ such that $|x_k-x|/\Delta < R$, $\Delta$ being the 
lattice spacing. For suitable $I(x,x_{k})$,
the function and its derivatives are evaluated to a
comparable accuracy \cite{dafref2}. This proves crucial in 
the evaluation of the disorder function.

The calculations presented are carried out using the
``Hermite DAF" \cite{dafref3}, defined by
\begin{equation}
I(x,x_k) = I(x-x_k)={\frac{\Delta}{\sigma}}{\frac{e^{-z^2}}{\sqrt{2\pi}}}
\sum_{j=0}^{M/2}\left( -\frac{1}{4}\right) ^{j}\frac{1}{j!}
H_{2j}(z),
\label{daf1}
\end{equation}
where $z=(x-x_k)/\sigma\sqrt{2}$
and $H_n(z)$ is the $n^{th}$ Hermite polynomial.
The Gaussian weight (of width $\sigma$)  in Eqn. (\ref{daf1})
makes $I(x_l -x_k)$ highly banded, reducing the computational
cost of applying the DAF to data.
The DAF representation of derivatives of a function known only
on a grid is given by
\begin{equation}
\left( \frac{d^{l}g}{dx^l}\right)_{DAF}(x)=\sum_{k}
\frac{d^l}{dx^l}I(x,x_k)g(x_k),
\label{daf2}
\end{equation}
which can be evaluated either on or off the grid.
In the continuum limit, the derivative of the DAF equals (exactly) the
DAF of the derivative \cite{dafref3}. 

The DAF approximation to a function that is sampled
on a square grid $(x_m,y_n)$ can be obtained using the two-dimensional
extension 
\begin{equation}
I((x,y),(x_m,y_n)) = I_X(x,x_m) I_Y(y,y_n) 
\label{2dDAF}
\end{equation}
of the approximating kernel \cite{zhuAhua}. Thus to estimate (say)
$\frac {d^2 v}{dx^2}$, Eqn. (\ref{daf2}) needs to be applied in
the $y$-direction (with $l=0$) and along the $x$-direction (with $l=2$).
(The application of the DAF operators in the two directions commute and can
be carried out in any order.)

As $M\rightarrow \infty$,
$I(x,x_k) \rightarrow \delta(x-x_k)$, and the DAF approximation
$g_{DAF}(x)$ becomes exact. With finite $M$, the sum
on the right side of Eqn. (3) becomes a polynomial (of order $M$),
resulting in $g_{DAF}(x)$ being smooth. Note that $g_{DAF}(x)$ can be
considered to be
a weighted running average of the signal. The Gaussian width
$\sigma$ determines the effect of neighboring points
in the DAF-approximation. The range $R$ is chosen to be sufficiently
large that the terms ignored in Eqn. (2) are negligible.
(The Gaussian weight included in the definition of $I(x,x_k)$
guarantees the decay of these terms.)

\section{Patterns Generated Using the Swift-Hohenberg equation}

The patterns analyzed in the paper are obtained from periodic fields 
$u({\bf x},t)$ generated by integrating
random initial states through a modified Swift-Hohenberg
equation (SHE) \cite{swiAhoh,croAhoh}
\begin{equation}
 \partial_t u  = D \Bigl(\epsilon - (k_0^2+\triangle)^2 \Bigr) u
               -\gamma u^3 - \nu (\nabla u)^2 .
\label{she}
\end{equation}
The parameters $D$, $k_0$, and $\gamma$ can be eliminated
through suitable rescaling of $t$, ${\bf x}$, and $u$ respectively.
$\epsilon$ measures the distance from the onset of patterns.
The results for the variational case ($\nu=0$) are presented in
this Section and those for the nonvariational case ($\nu\ne 0$) are
given in the next. 

The initial fields for the integration were random numbers in
a predetermined range. The time evolution is implemented using the 
Alternating Direction Implicit algorithm \cite{preAfla}. Each nonlinear term
$N\bigl [u({\bf x},t)\bigr ]$ is expanded to first order
in $\delta u = u({\bf x},t+\delta t)-u({\bf x},t)$, thus linearizing
the equations in $u({\bf x},t+\delta t)$. Updating the field involves
the inversion of a penta-diagonal matrix. The typical time step used for
the integration, $\Delta t \sim 0.1$, was chosen so that
the higher order terms in $\delta u$
are insignificant. We have confirmed the robustness of the 
integration by comparing (in a few cases) the results with those done 
for a smaller time-step ($\Delta t \sim 0.001$).

\subsection{Properties of the disorder function}

\begin{figure}
\narrowtext
\epsfxsize=3.0truein
\hskip 0.05truein
\epsffile{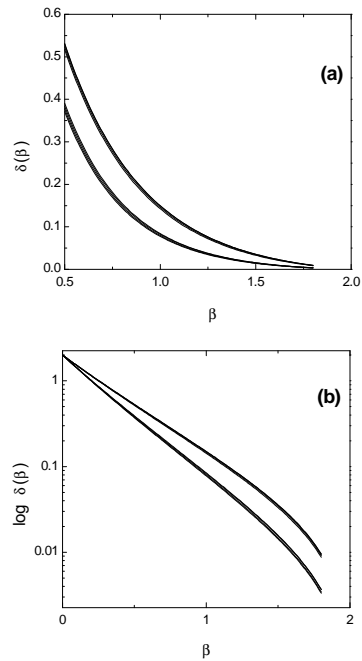}
\caption {The curves $\bar\delta(\beta)$ for patterns generated at 
two different sets of control parameters. The lower bunch consists of 
curves for four patterns at the first set of control parameters (Fig. 1) while
the upper bunch consist of those for a second set of
control parameters (Fig. 2). (b) shows the
same plots with a logarithmic vertical scale.}
\label{curves}
\end{figure}

The form of the measures $\bar\delta(\beta)$ were deduced by requiring
invariance under rigid motions of a single pattern. 
{\it Are these limited restrictions sufficient to yield characterizations that 
can delineate the observed ``commonality" in distinct patterns 
generated under identical conditions? }
Surprisingly, it appears to be the case. Fig. \ref{curves} shows
the disorder functions for several patterns. The curves
bunched at the bottom show $\bar\delta(\beta)$ for four 
textures (two of which are shown in Fig. \ref{set1})
generated at fixed control parameters. $\bar\delta(\beta)$ appears to
have captured the commonality of these distinct patterns.
Structures generated in the Gray-Scott model \cite{graAsco} and in a 
vibrated layer of
granular material \cite{melAumb} exhibit similar properties \cite{jone}.

The next question is if $\bar\delta(\beta)$
can differentiate between patterns with different visual characteristics.
Fig. \ref{set2} shows two structures obtained from the SHE
for a second set of control parameters. They have characteristics
that differ from patterns of Fig. \ref{set1}; e.g., they contain smaller 
domains and a larger density of defects. $\bar\delta(\beta)$ for four such 
textures are bunched together on the upper curves in Fig. \ref{curves}.
The significant separation of the two sets of curves
(e.g., the values of $\bar\delta(1)$ between the two sets
is about 25 times larger than the average difference between
curves within a set) confirms the ability of
$\bar\delta(\beta)$ to quantify the differences of the two groups of patterns.

\begin{figure}
\narrowtext
\epsfxsize=2.5truein
\hskip 0.3truein
\epsffile{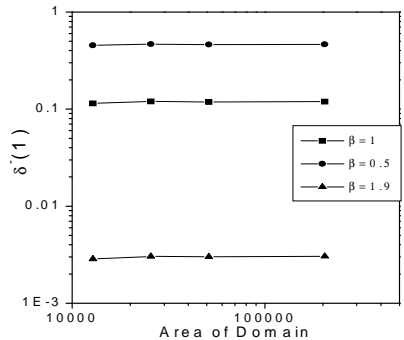}
\caption{The values of $\bar\delta(0.5)$, $\bar\delta(1.0)$ and 
$\bar\delta(1.9)$ for periodic patterns generated in domains of different
sizes. The areas of the domains are $36\pi\times 36\pi$,
$36\pi\times 72\pi$, $72\pi\times 72\pi$, and $144\pi\times 144\pi$.
In each case random initial states are evolved for 8000 time units 
under the SHE with control parameters given in Fig. 2.
The results indicate that $\bar\delta(\beta)$ are intensive variables
for labyrinthine patterns.}
\label{intensive}
\end{figure}

The disorder function quantifies the characteristics of
a labyrinthine pattern using the local curvature of the contour lines and
the wavelength variations, which 
typically increase with the (visual) disorder of a texture.
Thus, $\bar\delta(\beta)$ is able to quantify
the observation that patterns of Fig. \ref{set2} are more
disordered that those of Fig. \ref{set1}.

Next, we provide evidence to substantiate the claim that 
$\bar\delta(\beta)$ are intensive variables for labyrinthine patterns
such as those shown in Figs. \ref{set1} and \ref{set2} \cite{excep}. 
This is done by comparing values of $\bar\delta(\beta)$ for
patterns (with periodic boundary conditions) of several sizes.
The sizes of the domains chosen are $36\pi\times 36\pi$,
$36\pi\times 72\pi$, $72\pi\times 72\pi$, and $144\pi\times 144\pi$,
and each pattern is generated by integrating a random initial state
(with amplitude between $\pm 10^{-2}$) for a time $T=8000$ under the
SHE. The results, shown in Figure \ref{intensive}, give the mean
of 10 patterns for each domain size (except the largest
where only 5 patterns were used). The results indicate 
that $\bar\delta(0.5)$, $\bar\delta(1.0)$, and
$\bar\delta(1.9)$ are intensive variables, and the corresponding 
$\delta(\beta)$ are extensive variables.

\subsection{Relaxation of patterns}

The characterization of textures using 
$\bar\delta(\beta)$ finds one useful application in the study of the
relaxation from an initially random state.
Fig. \ref{snap-shots} shows several snapshots of a relaxing
pattern. During an initial period ($t<T_0\sim 800$) the local domains
emerge out of the random background and the mean intensity $<|u({\bf x},t)|>$
nearly reaches its final value. The subsequent evolution
due to domain coarsening is very
slow. These qualitative features are repeated in multiple runs
under the same control parameters.

Figure \ref{df_one} shows the behavior of $\Delta k$, 
$\bar\delta(0.5)$, $\bar\delta(1.0)$ and $\bar\delta(1.9)$ for the
evolution shown in Fig. \ref{snap-shots}. The curves remain
identical (except for small statistical fluctuations) for different
realizations of the experiment; i.e., the disorder
function captures configuration independent aspects of
the organization of patterns. The relaxation clearly
consists of two stages, with a sharp transition in $\bar\delta(\beta)$
at $t=T_0$ \cite{eldAvin,croAmei}. 
\end{multicols}

{\onecolumn
\begin{figure}
\caption{Several snap-shots of the relaxation of random initial state
whose intensity is between $\pm 10^{-2}$ under the SHE with 
$D=0.01$, $\epsilon=0.4$, $\nu=2$ and $k_0=1/3$. An initial phase
($t < 800$) when the local striped patterns are being formed
is followed by domain coarsening.} 
\label{snap-shots}
\end{figure}

\begin{figure}
\epsfxsize=5.5truein
\hskip 0.75truein
\epsffile{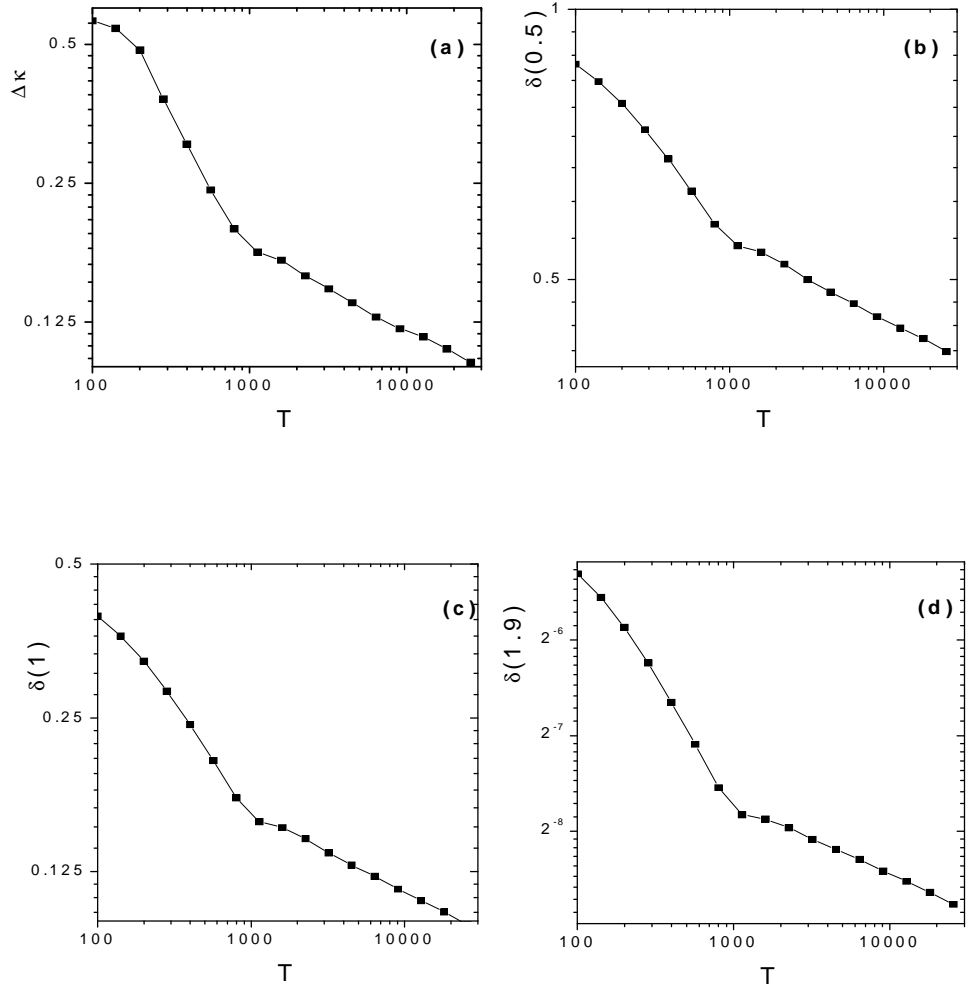}
\caption{The behavior of (a) $\Delta k$, (b) $\bar\delta(0.5)$ 
(c) $\bar\delta(1)$ and (d) $\bar\delta(1.9)$ during
the evolution shown in Fig. 6. During the initial phase $\bar\delta(\beta)$
decays (approximately) like $t^{-\frac{1}{2}\beta}$. During the 
second phase the scaling is nontrivial;
e.g., $\bar\delta(0.5) \sim t^{-0.09}$, $\bar\delta(1.0) \sim t^{-0.15}$
and $\bar\delta(1.9) \sim t^{-0.19}$. The transition between the
two phases occurs around $t=800$.}
\label{df_one}
\end{figure}}

\begin{multicols}{2}

During the initial phase, the 
time evolution of $\bar\delta(1)$ changes smoothly from a logarithmic
decay to a power law $\bar\delta(1) \sim t^{-\gamma_1}$, 
where $\gamma_1 \approx 0.5$. Corresponding $t^{-\frac{1}{2}}$
decay has been observed in the width of the structure factor \cite{eldAvin}.
The scaling is ``trivial" in the sense that for other ``moments"
$\bar\delta(\beta) \sim t^{-\frac{1}{2} \beta}$ \cite{footnote1}.
The decay of $\bar\delta(\beta)$ appears to be
associated with the $L \sim t^{\frac{1}{2}}$ growth of
domains in non-conserved systems \cite{rutAbra}.

The second phase of the relaxation (due to domain coarsening)
exhibits a more complex behavior. The moments $\bar\delta(0.5)$,
$\bar\delta(1)$ and $\bar\delta(1.9)$ behave approximately
as $t^{-0.09}$, $t^{-0.15}$ and $t^{-0.20}$ respectively,
indicating the presence of ``non-trivial" scaling \cite{chrAbra}.
Since the relative contribution of isolated defects increases with 
$\beta$ (Section II),
the slower decay of $\bar\delta(1.9)$ (compared to $\bar\delta(1)^{1.9}$)
suggests that changes in the density of defects is less
significant than the reduction of curvature of the contour lines \cite{eldAvin}.

\begin{figure}
\narrowtext
\epsfxsize=3.0truein
\hskip 0.05truein
\epsffile{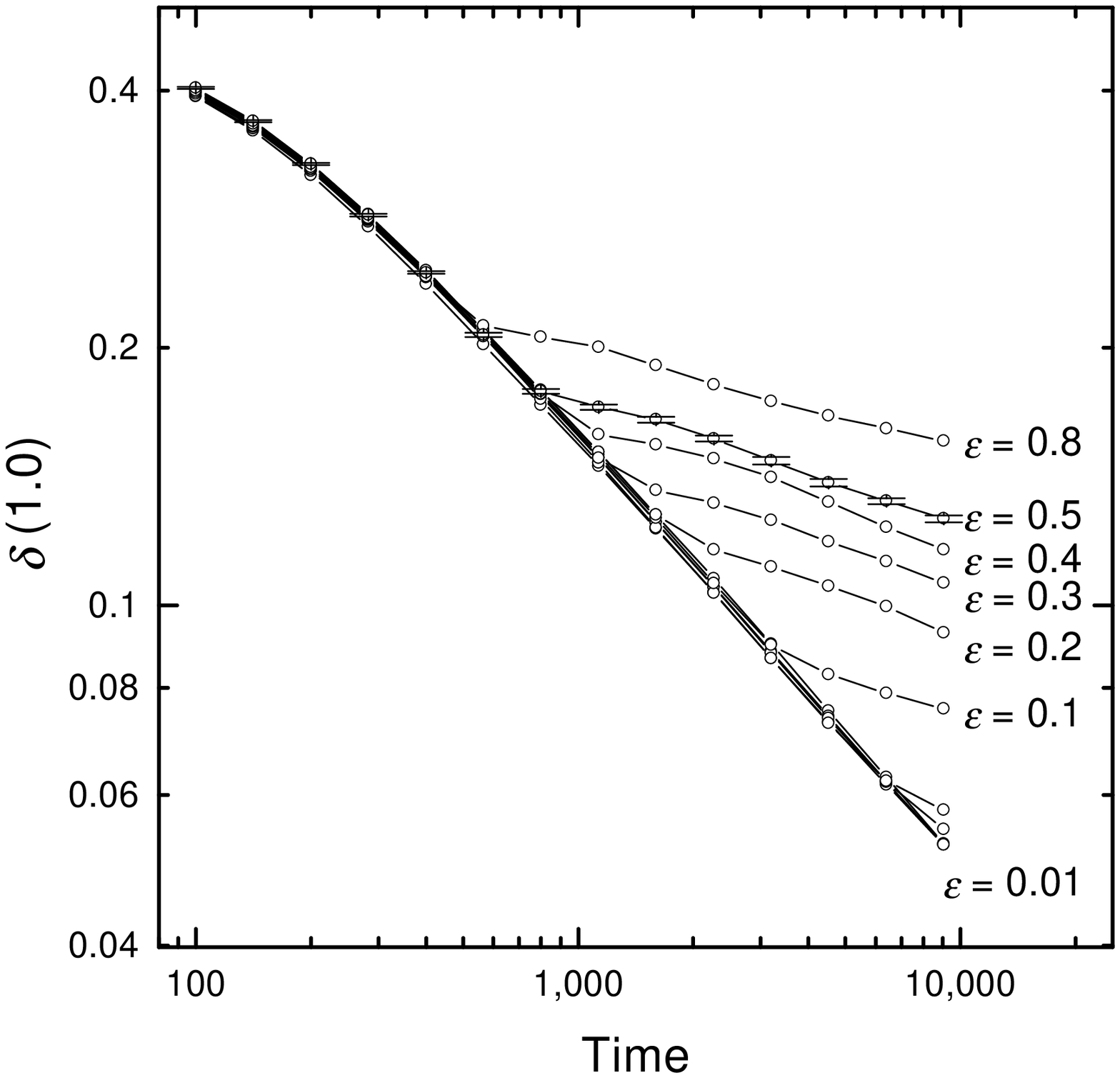}
\caption{The evolution of $\bar\delta(1)$ for patterns generated by 
integrating random initial states under the SHE for several values
of $\epsilon$. Each curve is an average of five runs. For clarity,
the standard deviations are shown only for one parameter value. 
For distinct $\epsilon$, $\bar\delta(1)$ exhibits identical behavior during 
initial growth of domains, and decays at the same rate during the coarsening
phase. The other moments $\bar\delta(\beta)$ exhibit similar behavior.}
\label{df_all}
\end{figure}

\subsection{Changes in the relaxation with $\epsilon$}

Figure \ref{df_all} shows the behavior of $\bar\delta(1)$ 
during the relaxation of random initial states
under the SHE for several values of $\epsilon$, all other
parameters being fixed. The initial decay of $\bar\delta(1)$
and the rate of decay during the second phase are seen to be 
independent of $\epsilon$. Furthermore, the transition between the
two phases advances with increasing $\epsilon$. 
Similar results are observed for all values of the moments $\beta$.

\begin{figure}
\narrowtext
\epsfxsize=3.0truein
\hskip 0.05truein
\epsffile{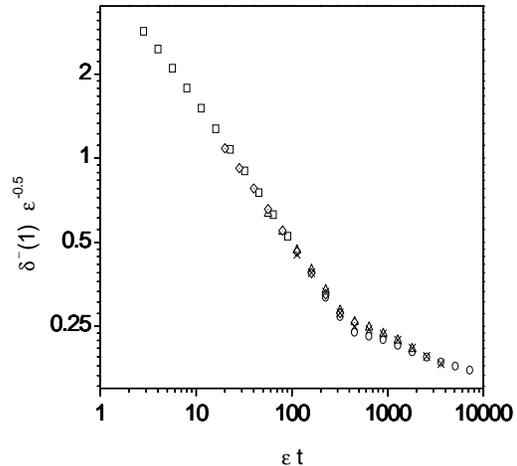}
\caption{The scaling function for the relaxation obtained by the rescaling
$t^{\prime} = t \epsilon$ and 
$\bar\delta^{\prime}=\bar\delta \epsilon^{-\frac{1}{2}}$. The data
correspond to $\epsilon$ values  of 0.01, 0.05, 0.1 0.4 and 0.8.}
\label{scale_all}
\end{figure}

Suitable scaling of variables, including 
$t \rightarrow \epsilon t$ can be used to eliminate $\epsilon$
from the SHE. Hence we expect that the rescaling 
$t \rightarrow \epsilon t$ and 
$\bar\delta(1) \rightarrow \epsilon^{-\frac{1}{2}} \bar\delta(1)$
will lead to collapse of the curves shown in Fig. \ref{df_all}.
This is indeed the case as seen from the scaling function 
(Fig. \ref{scale_all}).

\section{Relaxation in Nonvariational Systems}

In this Section we discuss properties of $\bar\delta(\beta)$ when
the spatio-temporal dynamics is nonvariational. 
The absence of an underlying ``energy" of the dynamics suggests
a faster relaxation, since the system cannot be constrained by
``metastable states" during the evolution.
The behavior of the disorder function confirms this expectation.

\begin{figure}
\narrowtext
\epsfxsize=3.0truein
\hskip 0.05truein
\epsffile{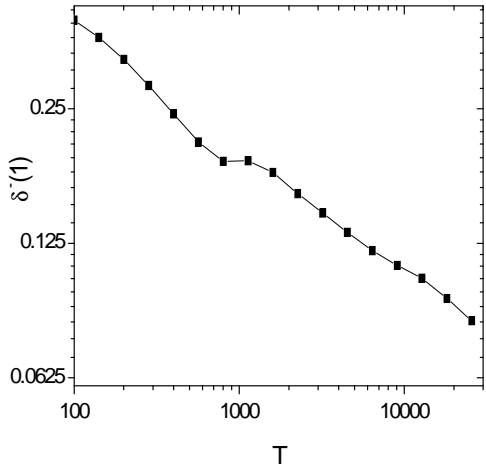}
\caption{The behavior of $\bar\delta(1)$ for the evolution of a random initial
state under the nonvariational modification of the SHE. The parameters
of the SHE are the same as given in Fig. 6 except for $\nu=2.0$. The 
initial decay is identical to the variational case while the 
coarsening phase exhibits a faster decay of $\bar\delta(1)$. }
\label{nv_one}
\end{figure}

\begin{figure}
\narrowtext
\epsfxsize=3.0truein
\hskip 0.05truein
\epsffile{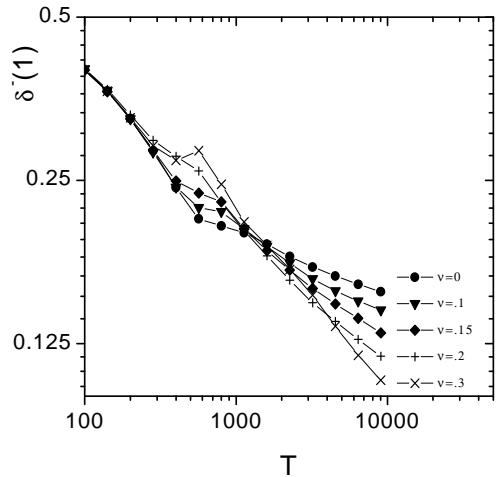}
\caption{The behavior of $\bar\delta(1)$ for several values $\nu$. Observe
that the decay during the growth phase remains the same, while the
decay during the coarsening phase increases with $\nu$.}
\label{nv_all}
\end{figure}

Figure \ref{nv_one} shows the behavior of $\bar\delta(1)$ for the
organization of a random field under a nonvariational SHE (i.e., $\nu \ne 0$).
The decay of $\bar\delta(1)$ remains the same (as the analogous
variational dynamics) during the initial relaxation 
(Fig. \ref{df_one}) and becomes faster during the coarsening phase.
As the coefficient $\nu$
of the ``nonvariational term" in Eqn. (\ref{she}) increases 
(the value of the remaining coefficients remaining the same), so does the 
relaxation rate during the coarsening phase (Fig. \ref{nv_all}).

The wave-number (obtained by minimizing 
$\bar\delta(1)$ over $k_0$) relaxes to a value ($k_0=0.61$) that is
larger than the corresponding one for the variational 
case ($k_0=0.59$). Such a deviation was observed earlier in
Ref. \cite{croAmei} where it was suggested that $k_0$ (for the
nonvariational case) corresponds to
the zero-climb velocity of isolated dislocation defects. 

\section{Discussion}

We have used the disorder function $\bar\delta(\beta)$ to characterize 
properties of textured patterns and their 
relaxation from initially random states. 
The disorder function was defined by requiring its invariance under 
rigid motions of a single texture. It was found to be
identical for multiple patterns generated under similar
external conditions; i.e., $\bar\delta(\beta)$ is configuration
independent. We provided evidence to confirm that the moments are
intensive variables. In addition, the disorder function
can differentiate between patterns with distinct characteristics.

The evolution of initially random states under the Swift-Hohenberg
equation is conveniently described using $\bar\delta(\beta)$.
The relaxation consists of two distinct stages separated
by a sharp transition. During the initial phase, local striped domains 
emerge out of the noisy background and their amplitudes saturate
close to their final value. This behavior is described by
a logarithmic decay followed by a power 
law decay of the disorder $\bar\delta(1) \sim t^{-\frac{1}{2}}$.
The scaling is ``trivial" in the sense that the decay of the 
remaining moments satisfy $\bar\delta(\beta) \sim \bar\delta(1)^{\beta}$. 
The second phase of the relaxation corresponds to domain
coarsening and is a much slower process. The scaling during this
phase is nontrivial.

As the system is driven further from the onset of patterns (as measured
by the parameter $\epsilon$) the duration of the initial phase
is reduced. However, the rates of decay of the disorder
function for the two phases 
remain unchanged and rescaling of time by $\epsilon$ and 
of $\bar\delta(\beta)$ by $\epsilon^{-\frac{1}{2} \beta}$ leads to
a scaling collapse. 

The addition of nonvariational terms to the spatio-temporal
dynamics leads to several interesting observations. The decay of
disorder during the initial phase is unchanged, and 
appears to be a model independent feature. Thus, one may expect to
observe it during relaxation of patterns
in experimental systems. The expectation of a faster relaxation
in nonvariational systems (due to the absence of ``potential 
minima") is seen only during domain coarsening.
This rate of relaxation is system dependent and increases as
the coefficient of the nonvariational term. 

There is very little theoretical understanding of the observed 
behavior of the disorder function. The decay of $\bar\delta(\beta)$ 
during the initial phase of pattern relaxation is reminiscent of 
analogous behavior in the XY-model \cite{kawa,lofAdeg}, 
and appears to correspond to the $L \sim t^{\frac{1}{2}}$ growth of
domains in nonconserved systems \cite{rutAbra}.

We would like to thank K. Bassler, D. K. Hoffman, R. E. Jones, D. J. Kouri and
H. L. Swinney for many stimulating discussions. This work was
partially funded by the Office of Naval Research  and the Energy Laboratory 
at the University of Houston.

\end{multicols}

\begin{thebibliography}{99}
\bibitem{footnote2} In general $\Delta k$ can be defined to be the distance
  between $k$-values for which $\bar\delta(1)$ is a factor $F\ (>1)$ of 
  the minimum. Properties studied in the paper, such as the decay rates,
  are independent of $F$.
\bibitem{excep} The disorder function is not intensive for 
  patterns with global regularity, such as a single target pattern
  or a single domain wall.
\bibitem{footnote1} The behavior during the entire first stage can be 
  approximated by $\bar\delta(\beta,t) = 1/(a+b log(t) + c t^{\beta/2})$, 
  for fixed $a$, $b$, and $c$.
\end{thebibliography}
\end{document}